\documentclass[3p,twocolumn]{elsarticle}

\usepackage{amssymb}
\usepackage{amsmath}

\usepackage{dblfloatfix}
\usepackage[hyphens]{url}
\usepackage[draft]{hyperref}
\usepackage{breakurl}

\usepackage{lineno}
\usepackage{xcolor}
\definecolor{darkgreen}{RGB}{0,100,0} 
\usepackage{float}
\usepackage{graphicx}
\usepackage{caption}
\usepackage{longtable}
\usepackage{placeins}
\usepackage{geometry}

\usepackage{algorithm}
\usepackage{algorithmic}
\usepackage[normalem]{ulem}
\usepackage{xcolor}

\newif\ifclean
\cleantrue   % ← PREPRINT / CLEAN
\usepackage[normalem]{ulem}
\usepackage{xcolor}

\ifclean
    \usepackage[final]{changes}
\else
    \usepackage[
      addedmarkup=colored,
      deletedmarkup=\sout
    ]{changes}

    \definecolor{addedcolor}{rgb}{0,0,1} % blue additions

    \renewcommand{\deleted}[1]{\textcolor{black}{\sout{#1}}}
\fi

\nolinenumbers

\begin{document}

\begin{frontmatter}


\title{Diffractive neural networks for mode-sorting with flexible detection regions}

\author[label1]{Kaden Bearne\corref{cor1}}
\ead{kaden.bearne@physics.ox.ac.uk} 
\author[label1]{Alexander Duplinskiy}
\author[label1]{ Matthew J. Filipovich}
\author[label1]{A. I. Lvovsky}

\affiliation[label1]{organization={Department of Physics, University of Oxford},%Department and Organization
            city={Oxford},
            postcode={OX1 3PU}, 
            country={UK}}

%% Corresponding author info
\cortext[cor1]{Corresponding author}

%% Abstract
\begin{abstract}
Mode-sorting is a procedure that decomposes a light field into a basis of transverse modes, directing each mode into a separate spatial location, allowing the constituent mode intensities to be measured simultaneously. We demonstrate a mode-sorter based on a diffractive optical neural network and show that it is advantageous to include the output detection regions into the trainable set of parameters of that network. This approach outperforms traditional mode-sorting methods, achieving \deleted{higher efficiency for the same crosstalk levels} \added{lower crosstalk levels for the same efficiency. For example, in sorting 25 Hermite-Gaussian modes with a 3 plate sorter, at 12 \% efficiency, the experimentally measured crosstalk decreases from  37.5 \% for fixed detection to 8.7 \% for flexible detection.} 
\end{abstract}

% Keywords

\end{frontmatter}

%%%%%%%%%%%%%%%%%%%%%%%%%%  body  %%%%%%%%%%%%%%%%%%%%%%%%%%
\section{Introduction}

Light is an excellent medium for information as it is fast and has various degrees of freedom, such as spectral, temporal, polarization, etc. Encoding information in transverse light structure is particularly beneficial for free-space communication, as it is robust to losses, dispersion \cite{Zhu2021vectorFS} and turbulence \cite{Krenn2014} and has a large information transfer capacity. While structuring beams arbitrarily is relatively straightforward using spatial light modulators (SLM) \cite{Bolduc:13}, demultiplexing a transverse light field into a given spatial mode basis, known as mode-sorting, has proven to be challenging.  Mode-sorting has applications in communication \cite{Puttnam2021}, imaging \cite{Tsang2016}, endoscopy \cite{Butaite2022} and optical machine learning \cite{Fang2024OAMML}. Mode-sorting is also an integral part of the spatial demultiplexing passive superresolution technique \cite{Dutton2019SRSaikat,Pushkina2021, Frank:23,Duplinskiy2025}.

Initial mode-sorters relied on transformations performed by standard optical components. For example, this approach  permits sorting Laguerre-Gaussian (LG) modes. These modes carry both a radial order and an orbital angular momentum (OAM). OAM is associated with a helical phase structure, which can be converted into a transverse phase ramp using the log-polar transformation. When subsequently focused by a lens, beams with different OAM will focus to different spatial locations. However, these locations overlap for neighbouring OAM eigenvalues, resulting in a $\sim 20$\% crosstalk \cite{Berkhout2010OAMsorting}.  Complementary to this, the radial order sorting is achievable using a fractional Fourier transform. The experimental demonstration sorting 3 different radial orders had a mean crosstalk of 15 \% \cite{Zhou2017radialLG}. Combining these two approaches allows for full sorting of the LG modal basis  \cite{Fu2018LGsorter}, with a crosstalk of 15.3 \% for a 10-mode sorter.

LG modes have a one-to-one relationship with the Hermite-Gaussian (HG) modes and can be converted using a pair of cylindrical lenses, enabling HG mode-sorting using an LG mode-sorter \cite{Zhou2018HGsorter}.  Both of these approaches rely on a fractional Fourier transformation, which limits the modal separation to two radial orders. It therefore becomes difficult to sort larger numbers of modes as these operations must be cascaded to perform the appropriate transforms. In addition, mode-sorting approaches based on a mode's distinct mathematical properties lack universality: currently known techniques are limited to a few bases such as LG or HG. 

A newer family of approaches involves multi-plane light converters (MPLCs), offering lower crosstalk for sorting the same number of modes. MPLCs are constructed using a series of programmable, spatially-variable phase plates separated by free-space propagation to implement a customizable transformation of a given input field. Originally proposed in 2010 \cite{Morizur2010UPMC}, this approach has been extended to mode multiplexing \cite{Labroille2014MPLC} and demultiplexing (mode-sorting) where the phase plates are reflections from an SLM. Fontaine {\it et al.} sorted 210 HG modes into  individual optical fibres using an MPLC with 7 phase plates with a crosstalk of 19\% \cite{Fontaine2019}. This approach has since been extended to other modal bases such as LG, OAM, Zernike, and even arbitrary speckles \cite{Kupianskyi2023}.  %The Zernike basis is of particular interest as it does not have a simple solution as the LG and HG basis do. 
For example, using a 5-plate sorter, up to 10 Zernike modes were separated with a 9.4\% crosstalk or up to 36 modes with a 31.2\% crosstalk \cite{Kupianskyi2023}. %MPLCs perform well They also show a clear promise in greatly increasing the number of modes that can be sorted as well as the types of modes that can be sorted.  
Additionally, the condition of orthogonality can be relaxed and overlapping quantum states can be sorted at the expense of introducing a loss \cite{Goel2023MPLCQuantumStates}. MPLCs can be taken to the limit of a single plane, acting as a modal beam-splitter sorting several modes with modest crosstalk \cite{Mazilu2017ModalBS}, or alternatively, the limit of infinite planes in the case of 3D graded-index volumes \cite{Barr20223DMPLC}. Fabricating a 3D structure, however, poses an additional challenge in a practical experiment.
Much effort has gone into characterizing \cite{Boucher2021MPLCcharacterization, Labroille2017Characterization} and optimizing \cite{Fang2021MPLCoptimization} the performance of MPLCs. %Unlike the physics-based mode-sorters the MPLC consists of arbitrary phase patterns designed to perform the desired transformation. 
Optimizing the phase patterns is a complex task, traditionally performed using an adjoint optimization algorithm called wavefront matching method (WMM) \cite{Hashimoto:05}. Training via gradient descent using various cost functions has also been used in simulation for mode-sorting, but no significant advantage with respect to wavefront matching has been observed \cite{Kupianskyi2023}. Other methods such as genetic algorithms have also been attempted \cite{Fickler2020GeneticAlgoritm}. % but there is not a definitive answer as to how to best design the phase patterns or even what the best transformation (i.e. desired output field) for mode-sorting would be. \\

In parallel, MPLCs have been explored by the optical machine learning community under the name of diffractive optical neural networks (DONNs). Since their initial introduction in 2018 \cite{Lin2018OzcanMNIST}, DONNs have been used for a variety of applications including deep learning, image recognition and reconstruction and communications \cite{sun2023,chen2024}. The physics of DONNs is identical to that of MPLCs, but DONNs are typically trained via backpropagation (gradient descent) on a digital twin. Hashimoto {\it et al.} argued that WMM can in fact be interpreted as a variant of gradient descent training, in which every optimization step increases the inner product between the forward and backpropagating modes, as we discuss in detail below \cite{Hashimoto2021WMMasDNN}. However, the backpropagation method appears to streamline the training of the MPLC, as shown in computational works  by Huang {\it et al.} \cite{Huang2021OAMsortingSim} and Zhu {\it et al.} \cite{zhu2023physicalNNtraining} and the experiment by Liu {\it et al.}, in which OAM beams have been (de)multiplexed \cite{Liu2023NNOAMsorting}.

Here we demonstrate a novel training method for mode-sorters using neural networks and flexible detection regions. We treat the mode-sorter as a DONN and train the phase plates to direct each mode into a separate detection region via backpropagation. The output detection regions are a part of the DONN's  trainable set of parameters. As a result, we achieve mode-sorting with significantly higher efficiencies (probabilities for the input photon in each input mode to reach the appropriate detection region) compared to existing methods, while maintaining similar levels of crosstalk. 

\section{Phase Plate Training Methods}

We begin by briefly describing existing methods of training MPLCs. 
An MPLC consists of phase plates separated by free-space propagation for the purpose of transforming the input electric field $\Psi_{\text{in}}(x,y)$ into a desired output field $\Psi_{\text{out}}(x,y)$. WMM computes the ``forward" optical field $\Psi_{p}^{\mathrm{for}}(x,y)$ as it propagates through each plate (indexed by $p$) from $\Psi_{\text{in}}$ at the input. Additionally, backward propagation $\Psi_{p}^{\mathrm{back}}(x,y)$ of the field, starting from $ \Psi_{\text{out}} $ at the output, is calculated. At each plate, the phase shift $\varphi_p(x,y)$ imposed by that plate is updated according to the phase difference between the forward and backward propagating fields\footnote{
In earlier versions of WMM, the phase was instead updated by a constant learning rate $a$ in the direction defined by the sign of the phase difference between the forward and backward propagating fields \cite{Sakamaki2007WMM}:
$$
\varphi_{p} \to \varphi_{p} - a\, \mathrm{sign}[\mathrm{Im}\{\Psi_{p}^{\mathrm{for}} [\Psi_{p}^{\mathrm{back}}]^{*}\}.
\label{eq:WMMupdate}
$$
} \cite{Fontaine2019, Fang2022OAMSorting}
\begin{equation}
\varphi_{p} \to \varphi_{p}- \mathrm{arg}\{\Psi_{p}^{\mathrm{for}} [\Psi_{p}^{\mathrm{back}}]^{*}\}.
\label{eq:WMMFontaine}
\end{equation}
The algorithm updates the phases according to Eq.~\eqref{eq:WMMupdate} at each plate for each pass, iterating through until convergence. We note that this update rule can be interpreted as a more general form of the single plate Gerchberg-Saxton phase retrieval algorithm \cite{Gerchberg:71}.
 
The WMM can be thought of as an algorithm attempting to maximize the overlap between the forward propagated output field and the desired output \cite{Sakamaki2007WMM,Kupianskyi2023}
 \begin{equation}
\mathrm{Loss} =- \left|\int\Psi_{\mathrm{out}}^{*}\Psi_\textrm{output plane}^{\mathrm{for}}{\rm d}x{\rm d}y\right|^{2}.
\label{eq:WMMrewardfunction}
\end{equation}
For the case of sorting spatially overlapping modes into spatially separate modes, the WMM iterations can be applied to each of the $n$ modes one-by-one, in which case the objective function becomes
\begin{equation}
\mathrm{Loss} = -\sum_{i = 1}^{n}\left|\int[\Psi_{\mathrm{out}}^{(i)}]^{*}\Psi^{\mathrm{for},(i)}{\rm d}x{\rm d}y\right|^{2}.
\label{eq:WMMrewardfunctionrelaxed}
\end{equation}
Importantly, updates \eqref{eq:WMMFontaine} are not precisely collinear with steepest descent direction with respect to the loss functions \eqref{eq:WMMrewardfunction} or \eqref{eq:WMMrewardfunctionrelaxed}, explaining better performance of gradient descent training via backpropagation \cite{Kupianskyi2023,zhu2023physicalNNtraining,Liu2023NNOAMsorting}.

Both WMM and the existing backpropagation implementations prescribe the exact mode of the output field. This is justified if the sorted modes need to be e.g.~coupled into single-mode fibers. However, this is often not necessary, ~e.g.~when the goal is only to determine the intensity of each mode in the input field. In this case, we need not prescribe the exact shape of the output modes, but only make sure that they land in different spatial regions of the output plane. 
This is the approach we take here. In addition to optimizing the phase plates in the traditional backpropagation manner, the training algorithm \added{(detailed in~\ref{appendix:Pseudocode})} also chooses a set $\{D_{j}\}$ of non-overlapping regions in the detection plane into which each mode is sent. \added{Each detection region can be either contiguous or consist of several sub-regions.} We find this innovation to significantly improve the performance of the mode-sorter. % These detection regions can be thought of as labeling each point in the detection plane as belonging to a given mode. So the network must optimize the phases at each plate as well as the detection regions for each mode . 

The performance of the system can be described as a matrix $I_{ij}$ of the total intensity that is found in the given output detection region $D_j$ when the input field is prepared in mode $i$:
\begin{equation}
I_{ij} = \int_{D_{j}}\left|\Psi_\textrm{output plane}^{(i)}\right|^{2}{\rm d}x{\rm d}y,
\label{eq:Ifunctionflexible}
\end{equation}
\added{where we assume the input intensity for each $i$th input mode to be unity.}
 We strive to maximize the efficiency of correct classification, \added{henceforth referred to as simply ``efficiency''.} %$ = \sum{|E_i|}^2D_{j}$. 
\begin{equation}
-\mathrm{Loss}_{\text{eff}} = \frac{1}{n} \sum_{i=1}^{n}  I_{ii}.
\label{eq:eff}
\end{equation}
while minimizing the modal crosstalk --- the total relative intensity of wrongly classified light
% \begin{equation}
% Loss_{xtalk} = \frac{1}{n} \sum_{i=1}^{n} \left( 1 - \frac{I_{ii}}{\sum_j I_{ij})\right}
% \label{eq:xtalk}
% \end{equation}
\begin{equation}
\mathrm{Loss}_{\text{xtalk}} = \frac{1}{n} \sum_{i=1}^{n} \left( 1 - \frac{I_{ii}}{\sum_j I_{ij}} \right).
\label{eq:xtalk}
\end{equation}
The final loss function is defined using hyperparameter $\alpha$ as a weighted combination of the modal crosstalk loss and the efficiency loss. 
\begin{equation}
\mathrm{Loss} = \alpha \mathrm{Loss}_{\text{eff}} + \mathrm{Loss}_{\text{xtalk}}(1 - \alpha).
\label{eq:LossFunctionFlexible}
\end{equation}
By varying $\alpha$, one can opt for levels of efficiency or crosstalk that best suit one's needs. Additionally, terms can be added to the loss function to impose further constraints on the phase plates, e.g.~smoothness or bit depth of $\varphi(x,y)$. 

\section{Numerical Simulation and Experimental Demonstration} \label{sec:Experiment}

\subsection{Setup and Mode-sorter Design} \label{sec:Setup}

\begin{figure}[ht!]
\centering\includegraphics[width=7cm]{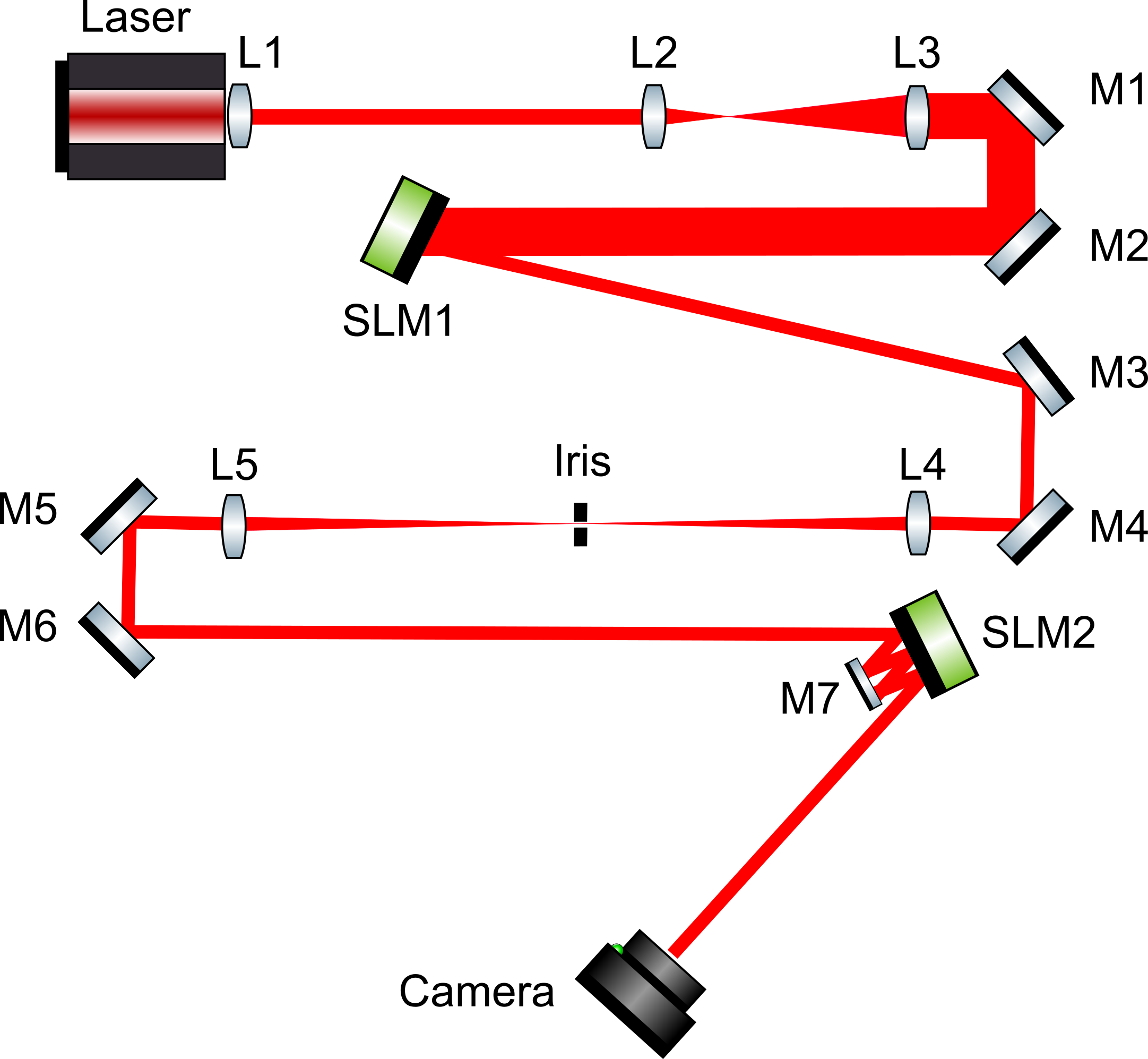}
\caption{Mode-generator and mode-sorting setup.}
\label{fig:Setup}
\end{figure}

Our mode-sorter and the setup for its characterization are shown in Fig.~\ref{fig:Setup} \added{(further details in~\ref{appendix:Setup})}. Initially, the laser (Toptica DL100) at 786 nm is coupled into a fibre to clean the beam, collimated at the output by L1. A telescope (L2,L3) serves to expand the beam so that it covers the entire active surface of SLM1 (Meadowlark 1920x1152). This SLM displays the hologram to generate arbitrary HG modes in the first diffraction order. Subsequently, the desired mode is imaged onto the MPLC mode-sorter by a 4f system (L4,L5) with an iris at its focal plane to select the first order. The MPLC is SLM2 (Meadowlark 1920x1152) facing parallel to a mirror (M7) to facilitate multiple reflections from different areas of the SLM, each acting as a phase plate. The mirror and the SLM are placed on translation and rotation stages to control the number of reflections as well as the distance between the plates. After the exit of the mode-sorter, a camera records the output intensity.  

To combat unmodulated light due to the reflection off the front SLM surface, we work in the first diffraction order and take care to prevent the unmodulated light from all plates from entering the final measurement. 

\subsection{Numerical Comparison of Training Methods}

\begin{figure*}[ht!]
% \centering\includegraphics[width=15cm]{Sorter-Draft/theoretical_w_inset_dB.png}
\centering\includegraphics[width=\textwidth]{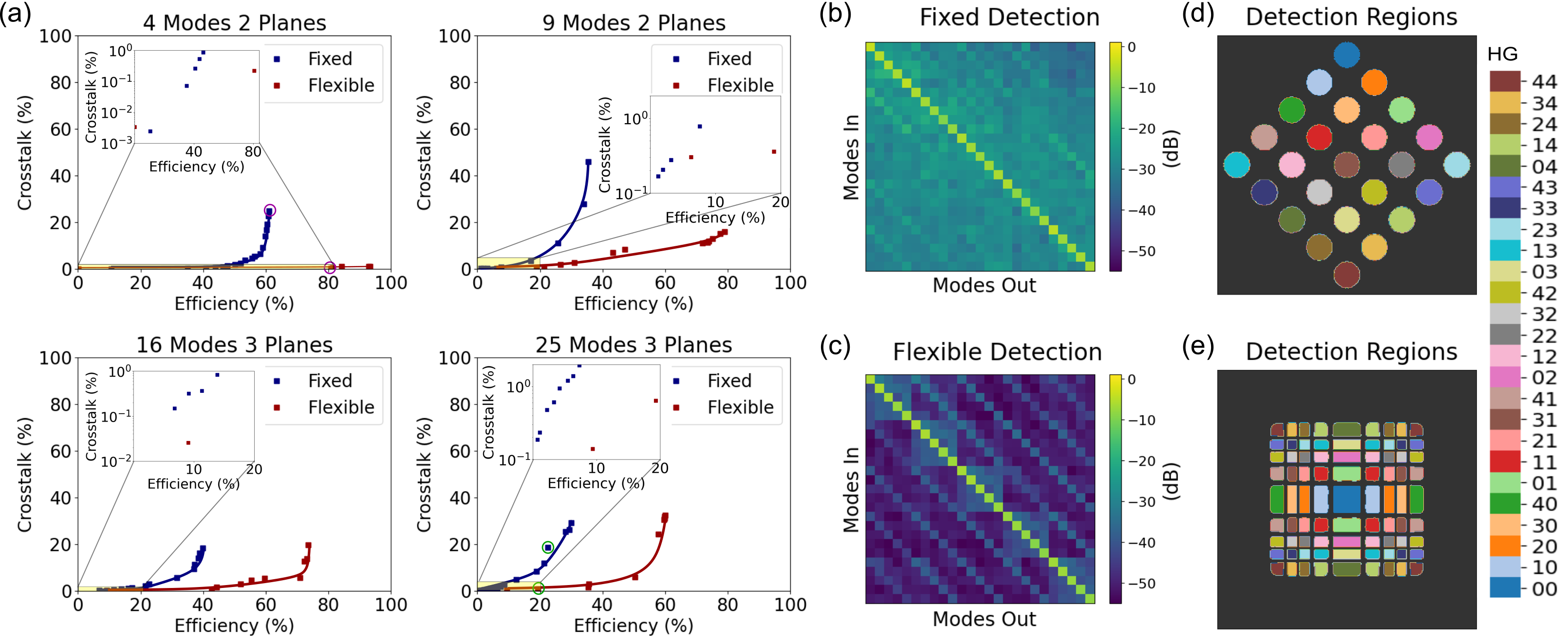}
\caption{(a) Crosstalk \eqref{eq:xtalk} vs. efficiency \eqref{eq:eff} (simulation) for mode-sorters trained with fixed \added{(d)} and flexible \added{(e)} detection regions. \added{(b, c) Crosstalk matrices \eqref{eq:Ifunctionflexible} and (d,e) detection regions for the 3-plate fixed (b,d) and flexible (c,e) mode-sorters for 25 HG modes at points shown by green circles in (a). The value in decibels corresponds to $10\log I_{ij}$ }  \deleted{The data for fixed detection regions were generated assuming these regions to be circular at fixed locations, with the radii and center positions chosen such that the total area is similar to the total area of the flexible detection regions.} The purple \deleted{and orange} circles show the points for which the data in  Fig.~\ref{fig:ExpDetectionRegions} are calculated.  %The loss function is given in Eq. \ref{eq:LossFunctionFlexible} and the crosstalk is computed as either the field overlap or intensity overlap depending on the case. 
}
\label{fig:Theoretical}
\end{figure*}

 We perform numerical simulations to train the phase plates and test their performance in various experimental situations. These simulations are performed in Python using TorchOptics \cite{TorchOptics2024}, a package for simulating and training free-space optical systems using the framework of PyTorch, leveraging GPUs and CUDA. Crosstalk \emph{vs} efficiency plots for sorting 25 modes (HG$_{00}$ to HG$_{44}$) with a 3-plate mode-sorter with the geometric parameters matching those of our experiment are shown in  Fig.~\ref{fig:Theoretical}, where the efficiency is defined as the negative of the right-hand side of Eq.~\eqref{eq:eff} and the crosstalk by Eq.~\eqref{eq:xtalk}. To determine the benefit of flexible detection, we train two different neural networks to find the phases, one with fixed detection and the other with flexible detection regions, using the loss function  Eq.~\eqref{eq:LossFunctionFlexible}.

 \added{Both the fixed and flexible detection regions were defined within the same output area. The fixed design uses a diamond arrangement of circles around Gaussian output spots, akin to previous works \cite{Kupianskyi2023, Fontaine2019}. %) scaled and spaced to match the area available to the flexible detection regions. 
 The fixed region sizes are chosen to balance efficiency and crosstalk, as detailed further in~\ref{appendix:SizingFairness}. %further optimizing their size or spacing for either metric would effectively turn the baseline into a tunable detection rather than a fixed one.
 }

When the hyperparameter $\alpha$ is varied, the network finds solutions with different levels of crosstalk and efficiency represented by the two curves in Fig.~\ref{fig:Theoretical}(a). For $\alpha = 1$ the system maximizes the efficiency and neglects the crosstalk, as represented by the final point (upper right) of the curves. Looking at the 25 mode 3-plate sorter, for the fixed detection regions, this yields a 30\% efficiency at a 29\% crosstalk. For flexible detection regions, the same crosstalk level (29\%) is achieved at a 58.5\% efficiency. 
%While one could take the approach of changing the size or spacing of the fixed output field in order to maximize efficiency, this would defeat the point of having a fixed output and raises more questions in terms of what may be optimal. Flexible detection regions naturally answer this by letting the network chose the detection regions. 
Alternatively, for $\alpha \approx 0$, shown in the inset to Fig.~\ref{fig:Theoretical}(a), the main goal of the network is preventing crosstalk. For all cases, we see the crosstalk levels to saturate below 1\%, with further crosstalk improvements being minimal with significant cost in efficiency. \added{As representative examples, the green circled points from Fig.~\ref{fig:Theoretical}(a) are presented for sorting 25-modes with 3 plates with fixed (b) and flexible (c) detection regions. These were obtained using $\alpha = 0.75$ for the fixed and $\alpha = 0.1$ for the flexible cases, resulting in a crosstalk of 18.7 \% and efficiency of 22.6 \% for the fixed, and 0.63 \% crosstalk and 19.3 \% efficiency for the flexible.}

\subsection{Experimental Results}

\begin{figure*}[ht!]
\centering\includegraphics[width=\textwidth, height=0.4\textheight, keepaspectratio]{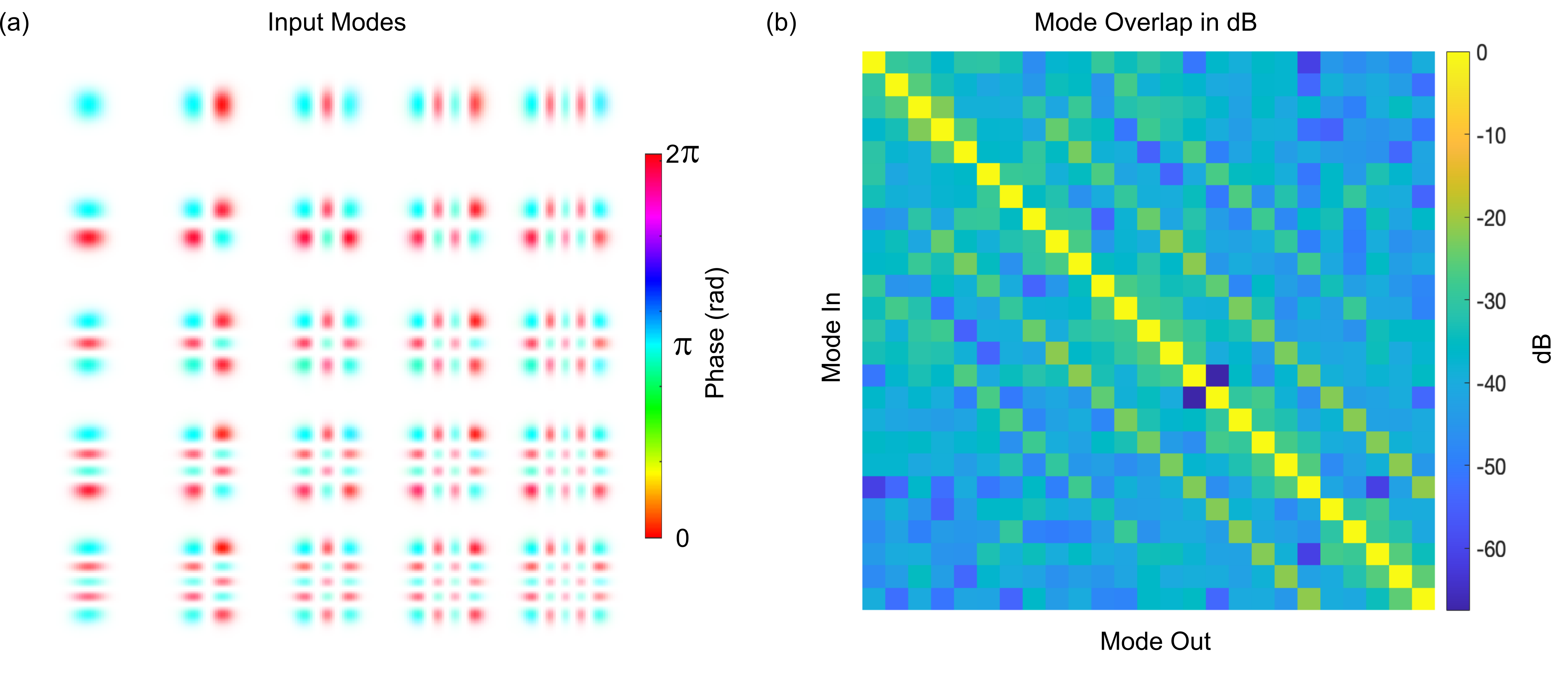}
\caption{The generated input modes characterized via off-axis holography (a) and their power overlap matrix (b.)% computed as $10\log_{10}(|HG_{in}^{*} HG_{out}|^2)$ in dB or as a percent $100 \times 10^{\text{dB}/10}$
}
\label{fig:Modes}
\end{figure*}

\begin{figure*}[hb!]
\centering\includegraphics[width=\textwidth, height=0.4\textheight, keepaspectratio]{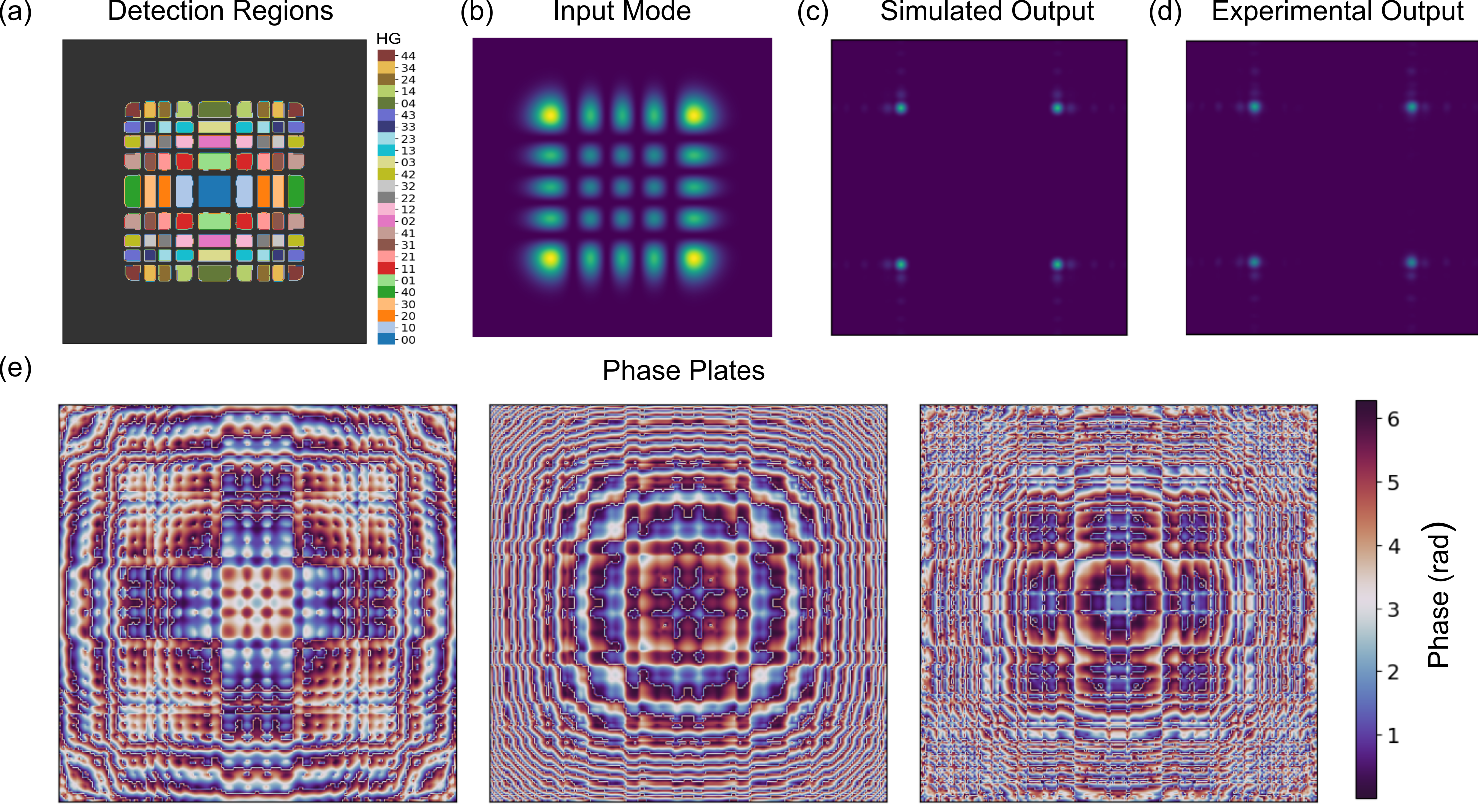}
\caption{3-plate, 25-mode sorter. (a) Flexible trained detection regions, different modes are represented by different colours. (b) Input mode $\mathrm{HG}_{44}$. (c) Simulated output for this mode. (d) Experimental output for this mode. (e) Trained phase plates.}
\label{fig:3planes25modes}
\end{figure*}

Before we can analyze the mode-sorter performance, we need to characterize the HG modes generated via SLM1. We solve this task using the method of Bolduc {\it et al.} \cite{Bolduc:13}, using off-axis holography \cite{Cuche2000Holography}. The reconstructed modes are shown in Fig.~\ref{fig:Modes}(a) along with their overlap matrix in Fig.~\ref{fig:Modes}(b). %The overlap is computed by taking the intensity of the complex overlap of the two fields. 
The modes were found to have a mean fidelity of 97.8 \% and a mean modal overlap of 1.5 \% for 25 modes. This represents the limit for the lowest achievable crosstalk after mode-sorting.

We demonstrate three versions of the mode-sorter: 1 plate for 4 modes, 2 plates for 4 and 9 modes, and 3 plates for 16 and 25 modes. The HG modes have a waist size of 29 pixels for an SLM pixel size of 9.2 $\mu$m. Each phase plate measures $200\times 200$ pixels. The separation between the SLM and the mirror is 4.25 cm. This spacing allows enough propagation distance to separate the diffraction orders and make full use of each phase plate.

% \begin{figure}[ht!]
% \centering\includegraphics[width=12cm]{detection-regions-4modes.png}
% \caption{Comparison of trained detection regions (colour coded) for sorting 4 modes with two planes corresponding to (a) purple and (c) orange circles in Fig.~\ref{fig:Theoretical} (a), and (b) purple and (d) orange circles in 
% Fig.~\ref{fig:Tradeoffplot} (a). }
% % a) simulation with efficiency of 81 \% and crosstalk of 0.22 \%, b) experiment with efficiency of 46 \% and crosstalk of 2.0 \% , c)  simulation with efficiency of 93 \% and crosstalk of 1.0 \%, and d) experiment with efficiency of 55 \% and crosstalk of 5.1 \%.}
% \label{fig:ExpDetectionRegions}
% \end{figure}

\begin{figure*}[ht!]
\centering\includegraphics[width=\textwidth, height=0.5\textheight, keepaspectratio]{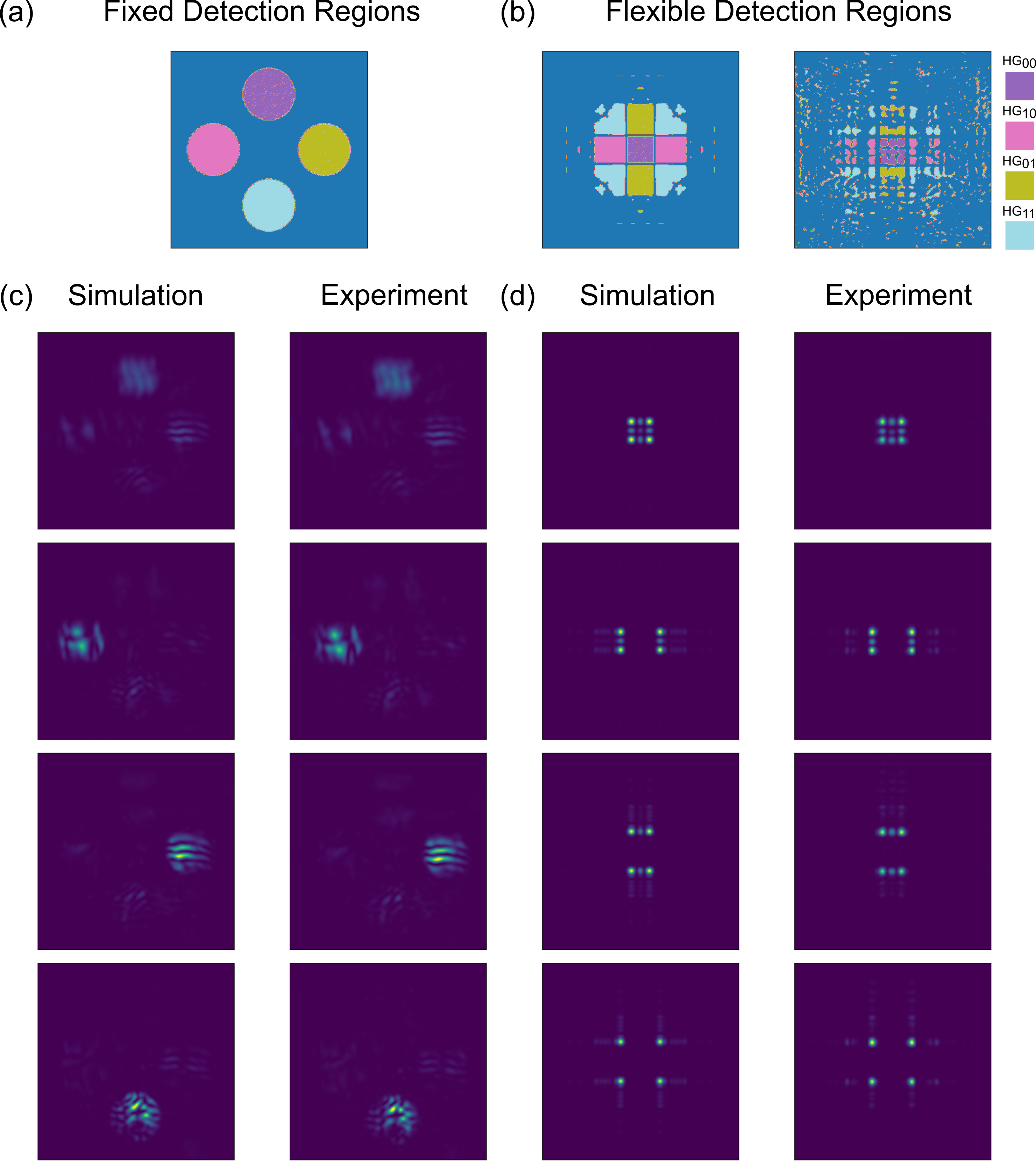}
\caption{\added{Comparison of (a) fixed and (b) flexible detection regions (colour coded) for sorting 4 modes with two plates corresponding to purple circles in Fig.~\ref{fig:Theoretical} (a) (left) and
Fig.~\ref{fig:Tradeoffplot} (a) (right). The resulting outputs for these cases are shown for the fixed (c) and flexible (d) detection regions; top to bottom: HG00, HG10, HG01, HG11. }}

\label{fig:ExpDetectionRegions}
\end{figure*}

Figure \ref{fig:3planes25modes} shows \deleted{the results for the} \added{the example of a flexible detection} 3-plate sorter \deleted{with} \added{for} 25 HG modes \added{whose crosstalk matrix is presented in Fig.~\ref{fig:Theoretical}}. \deleted{The training finds the detection regions [Fig.~\ref{fig:3planes25modes}(a)] and the phase plates [Fig.~\ref{fig:3planes25modes}(e)].} \added{The detection regions [Fig.~\ref{fig:3planes25modes}(a)] as well as the phase plates [Fig.~\ref{fig:3planes25modes}(e)] are learned parameters.} The \deleted{trained} phase plates are symmetric, reflecting the inherent symmetry of the HG mode set. While training, slight numerical instabilities can lead to asymmetric plates. For this reason, we force the plates to remain symmetrical through appropriate parametrization. This makes little difference in simulated performance but greatly aids in experimental alignment.

\begin{figure*}[ht!]
%one line
% \centering\includegraphics[width=15cm]{Sorter-Draft/experimental-all-tradeoff-2.png}
%two lines
\centering\includegraphics[width=\textwidth]{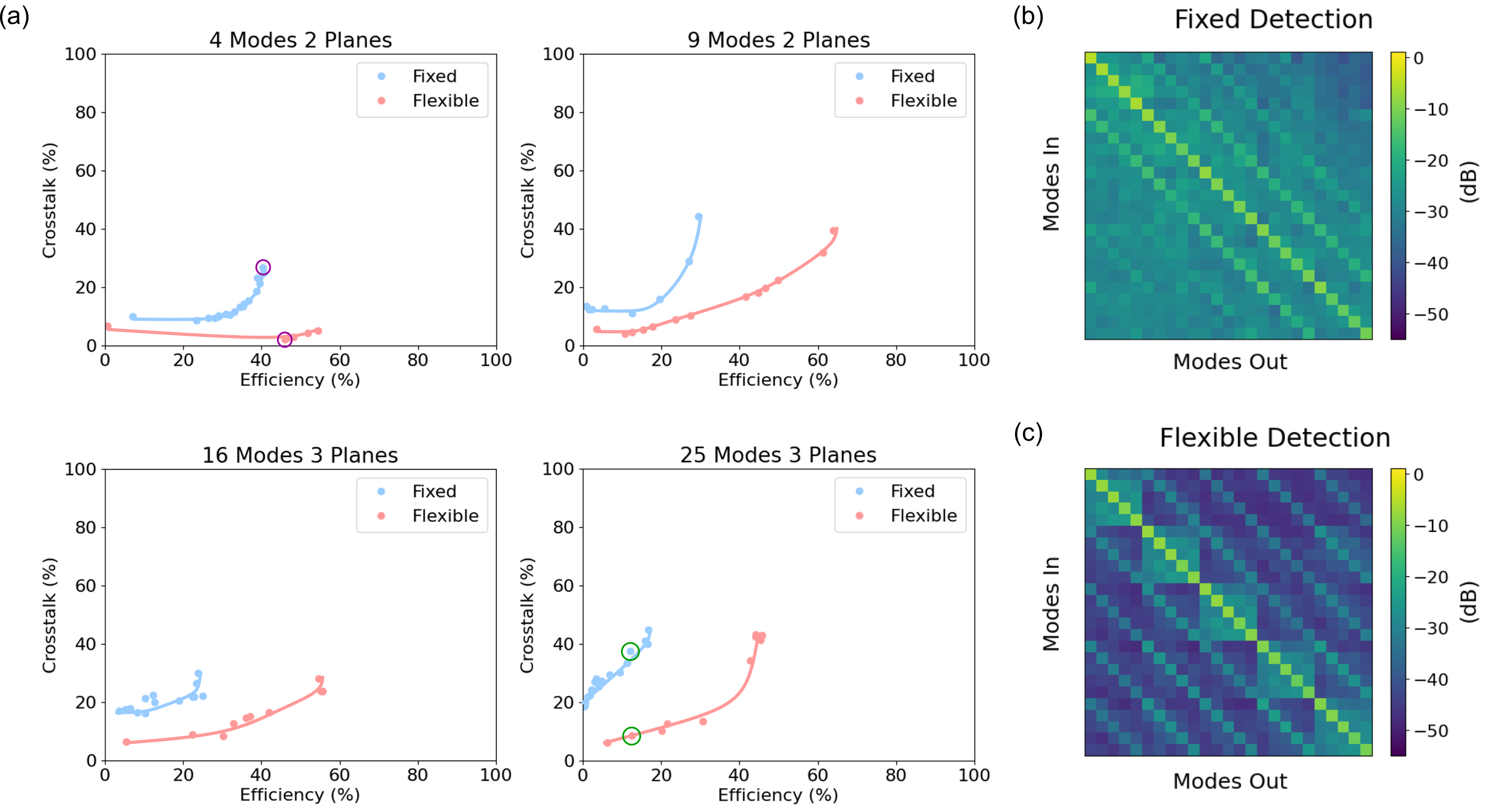}
\caption{(a) Crosstalk \eqref{eq:xtalk} vs. efficiency \eqref{eq:eff} (experimental), shown for mode-sorters trained with fixed and flexible detection regions. The efficiency is evaluated as the ratio of the power of a sorted mode output in the corresponding detection region and the output intensity of that mode with all phase plates set to $\varphi_p(x,y)\equiv0$. \deleted{(b) Crosstalk matrix for the 3-plate flexible mode-sorter for 25 HG modes (green circle).} \added{(b, c) Crosstalk matrices \eqref{eq:Ifunctionflexible} for the 3-plate fixed (b) and flexible (c) mode-sorters for 25 HG modes at points shown by green circles in (a). The value in decibels corresponds to $10\log I_{ij}$. } The purple \deleted{and orange} circles show the points for which the data in  Fig.~\ref{fig:ExpDetectionRegions} are calculated.}
\label{fig:Tradeoffplot}
\end{figure*}

%These are then applied to the SLM with a pixel precision using the flexible plates for alignment as they are symmetric. 
\added{It has been shown that randomly shifting the phase plates during training can make a DONN more robust to experimental misalignment for classification tasks \cite{Ozcan2020Vaccination}. A brief study of the effect of a phase plate shift is presented in~\ref{appendix:Alignment}. However, for the task of spatially sorting modes in experiment, we found the precision of plate alignment to be critical, with displacement by a fraction of an SLM pixel (<10 $\mu m $) significantly increasing crosstalk. To address this, we fine-tune the location of the beam by adjusting M5--7 and SLM2 (Fig.~\ref{fig:Setup}).} \deleted{We found the precision of the input light field location with respect to the  phase plates to be critical, with the displacement by a fraction of an SLM pixel significantly affecting the performance. To address this, we fine-tune this location  by adjusting M5-7 and SLM2. }

%While these plates have corresponding simulated detection regions, properly aligning the output sorted modes captured by the camera to these is challenging. Instead of this, we train detection regions for the flexible experimental data (pink circles in Fig. \ref{fig:Tradeoffplot}).  

\begin{figure}[ht!]
\centering\includegraphics[width=7cm]{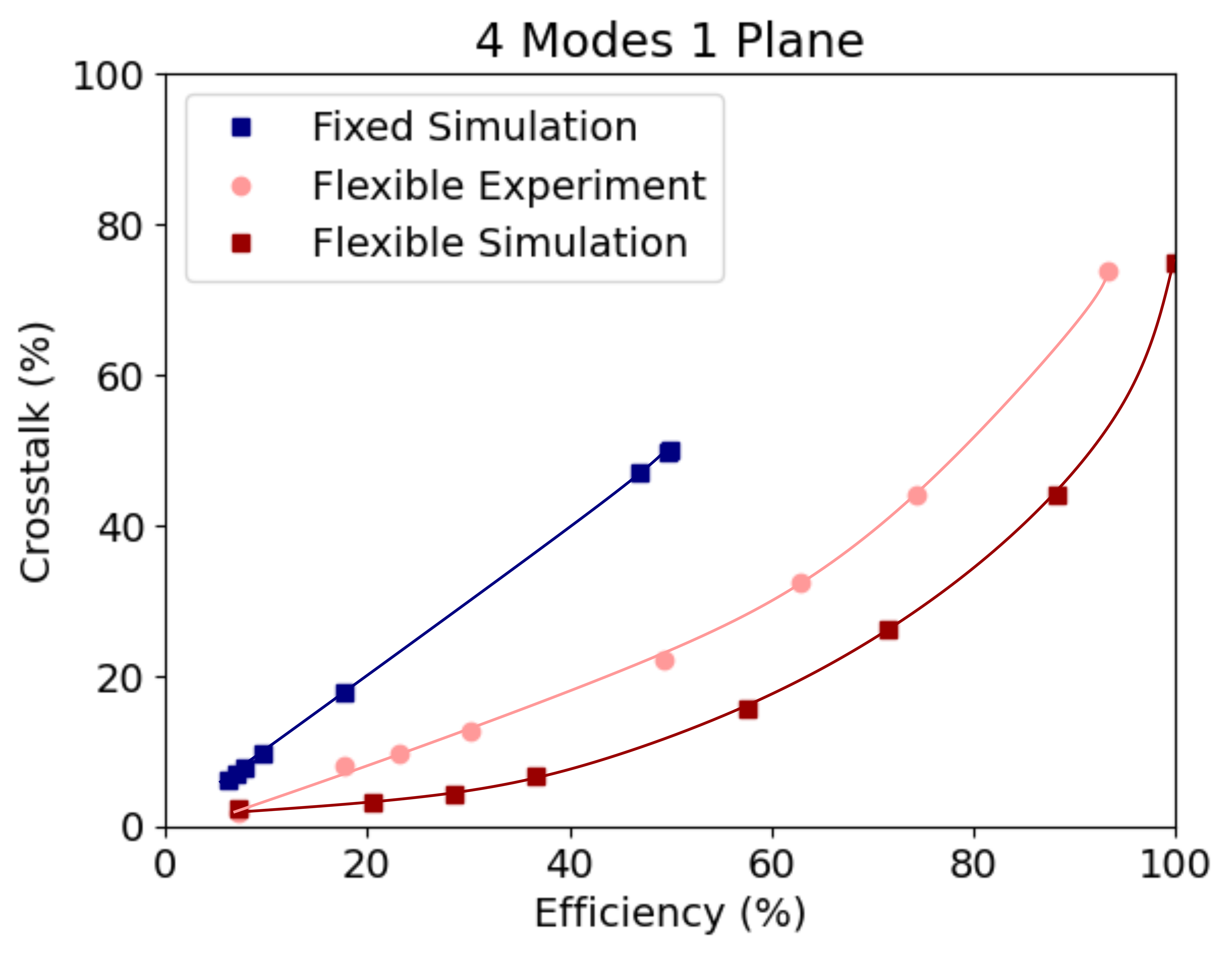}
\caption{Crosstalk vs. efficiency for a single-plate 4-mode sorter, shown for mode-sorters trained with fixed and flexible detection regions.}
\label{fig:TradeoffplotSinglePlane}
\end{figure}

Experimental imperfections cause the mode output fields to deviate from their theoretically predicted shapes. To address this, we re-optimize the detection regions accounting for the experimentally measured outputs. \deleted{ as illustrated in  
%The resulting regions are roughly the same as the simulated regions with noise. 
Fig.~\ref{fig:ExpDetectionRegions}.} \added{As an example, the case of a 2-plate 4 HG mode sorter, corresponding to the purple circles in  Fig.~\ref{fig:Theoretical} (a) and
Fig.~\ref{fig:Tradeoffplot} (a) is presented in Fig.~\ref{fig:ExpDetectionRegions}. }

%To properly measure efficiency, in simulation the input fields are normalized. 
The mode-sorter performance results are summarized in Fig.~\ref{fig:Tradeoffplot}. \added{These represent the experimental demonstrations of the simulated results from Fig.~\ref{fig:Theoretical}}. The mode-sorter with flexible detection regions outperforms its fixed detection counterpart. \added{The representative examples, the green circled points from Fig.~\ref{fig:Tradeoffplot} (a) are again presented for sorting 25-modes with 3 plates with fixed (b) and flexible (c) detection regions. For the same 12 \% efficiency, the crosstalk of 37.5 \% for the fixed decreases to 8.7 \% for the flexible.} While both methods offer a trade-off between the crosstalk and efficiency, the fixed regions appear to have an earlier saturation point where the efficiency can no longer be improved at the expense of crosstalk. 

We attribute most of the simulation-experiment gap to the various imperfections that come from using an SLM. \deleted{including the SLM cavity effect and pixel crosstalk \cite{Pushkina:20}.} \added{While SLM backplane curvatures and voltage look-up tables are individually calibrated, other sources of error include the SLM cavity effect, pixel crosstalk \cite{Pushkina:20}, and phase-quantization \cite{HarrisonSLMBitDepth}.} The resulting error accumulates with each phase plate. We believe that transmissive fabricated phase plates would not suffer from many of these issues \cite{Vesela2024PhasePlates}. 
 
Figure \ref{fig:TradeoffplotSinglePlane} shows the performance of a single-plate sorter with four modes. Remarkably, the flexible-detection mode-sorter performs in the experiment better than the fixed-region mode-sorter in simulation. This may be beneficial in many imaging applications such as phase-contrast \cite{Burch:42} or dark-field \cite{Gage:20} microscopy, where the goal is to prevent the light from the $\mathrm{HG}_{00}$ mode from contaminating other channels. % The flexible approach  achieves this efficiently with a single plate. 
%In addition to this, due to the ability to tradeoff between crosstalk and efficiency we might consider different potential applications. 

\section{Conclusion}
In conclusion, we have demonstrated that mode-sorters can be trained using a neural network approach, akin to DONNs, and that allowing for flexible detection regions outperforms the traditional MPLC fixed detection. Using the trade-off between crosstalk and efficiency, different applications (e.g. imaging or communication) can choose the desired levels for the respective tasks. For example, in a task requiring state discrimination such as optical communication using multiplexed  OAM states \cite{Yao:11}, the crosstalk can be quite high and so efficiency should be favoured.

The design for the physical mode-sorter is simple and requires only an SLM and a mirror, allowing for easy reproducibility in most optics labs. For applications where the mode-sorter need not be reconfigurable, phase plates can be fabricated to avoid the spurious SLM effects and give a better performance.

\section*{Acknowledgments}
The project is funded by EPSRC Standard Grant EP/Y020596/1 and EPSRC Impact Acceleration Account Award EP/X525777/1. KB is supported by the Clarendon Fund, Lincoln College Elman Poole Graduate Scholarship, and EPSRC DTP Studentship EP/W524311/1.

\section*{Disclosures}
The authors declare no conflicts of interest.

\section*{Data availability} Data underlying the results presented in this paper may be obtained from the authors upon reasonable request.
% %% The Appendices part is started with the command \appendix;
% %% appendix sections are then done as normal sections
% \appendix
% \section{Example Appendix Section}


% %% For citations use: 
% %%       \cite{<label>} ==> [1]

% %%
% Example citation, See \cite{lamport94}.

\bibliographystyle{elsarticle-num} 
\bibliography{main}

\appendix

\section{Training}
\label{appendix:Pseudocode}

\begin{algorithm}
\caption{Training procedure for DONN}
\begin{algorithmic}[1]
\STATE Choose $\alpha$ for training loss function \\
\STATE Initialize model of DONN \\
\STATE Initialize optimizer and learning rate \\[0.5em]
\FOR{each training iteration}
    \STATE Model in training mode
    \STATE Zero gradients
    \STATE Pass input field through model of DONN
    \STATE Get output from DONN
    \STATE Calculate Efficiency Loss according to Eq.~\eqref{eq:eff}
    \STATE Calculate Crosstalk according to Eq.~\eqref{eq:xtalk}
    \STATE Calculate Losses according to Eq.~\eqref{eq:LossFunctionFlexible}
    \STATE Compute Gradients
    \STATE Update Parameters
\ENDFOR
\end{algorithmic}
\end{algorithm}

\begin{algorithm}
\caption{DONN Model}
\begin{algorithmic}[1]

\STATE Start with input fields \\

\FOR{each phase plate}
    \STATE Apply phase plate (trainable parameter) to input fields
    \IF{not last plate}
        \STATE Propagate fields forward to next phase plate
    \ELSE
        \STATE Propagate fields to camera plane
    \ENDIF
\ENDFOR

\STATE Compute output intensities at the camera

\IF{Training flexible detection regions}
        \STATE Detection regions $D_j$ (trainable parameter) 
    \ELSE
        \STATE Use fixed detection regions$D_j$
    \ENDIF
\STATE Compare with detector targets using Eq. \eqref{eq:Ifunctionflexible}
\STATE Return output

\end{algorithmic}
\end{algorithm}

\added{
The detection regions $D_j$ are parameterized as trainable weight maps $w_j (x,y)$ over the detection plane, with $j$ ranging from zero to the number of regions. The value $j=0$ corresponds a pixel not being associated with any region. The intensities \eqref{eq:Ifunctionflexible} are calculated during the training according to 
$$I_{ij}=\sum_{x,y}\left|\Psi_\textrm{output plane}^{(i)}\right|^{2}\sigma(\mathbf{w}(x,y))_j,$$
where $\sigma(\mathbf{w}(x,y))_j$ is the softmax function over all elements of the vector $\mathbf{w} (x,y)$ at pixel $(x,y)$. From these intensities, the loss function \eqref{eq:LossFunctionFlexible} is computed. As the training progresses, one element of each vector $\mathbf{w} (x,y)$ becomes dominant, so each pixel becomes associated with a particular single detection region, or no detection region. 
}

\added{
An alternative strategy of training was to use a sigmoid transform instead of softmax. Training until convergence pushes each weight close to 0 or 1. This strategy slightly outperformed softmax-based training for low values of $\alpha$. However, for high values of $\alpha$, several elements of $\mathbf{w} (x,y)$ at a particular pixel $(x,y)$ could converge to 1, resulting in overlapping detection regions.  The regions trained using sigmoid were therefore checked for overlap, and if found, retrained using the softmax transform which strictly enforces the non-overlapping condition.}

% \textcolor{magenta}{\textit{explain why this automatically  does not happen normally if we are not doing extreme cases.}}

\added{In both fixed and flexible cases the model is trained for various values of $\alpha$ ranging from 0 to 1. The solutions lying on the Pareto-optimal frontier (trade-off between efficiency and crosstalk) were selected for further experimental analysis and displayed in Fig. \ref{fig:Theoretical} and Fig. \ref{fig:Tradeoffplot}.}

\section{Full Experimental Setup}
\label{appendix:Setup}

\begin{figure*}[t]
\centering\includegraphics[width=\textwidth,height=0.5\textheight, keepaspectratio]{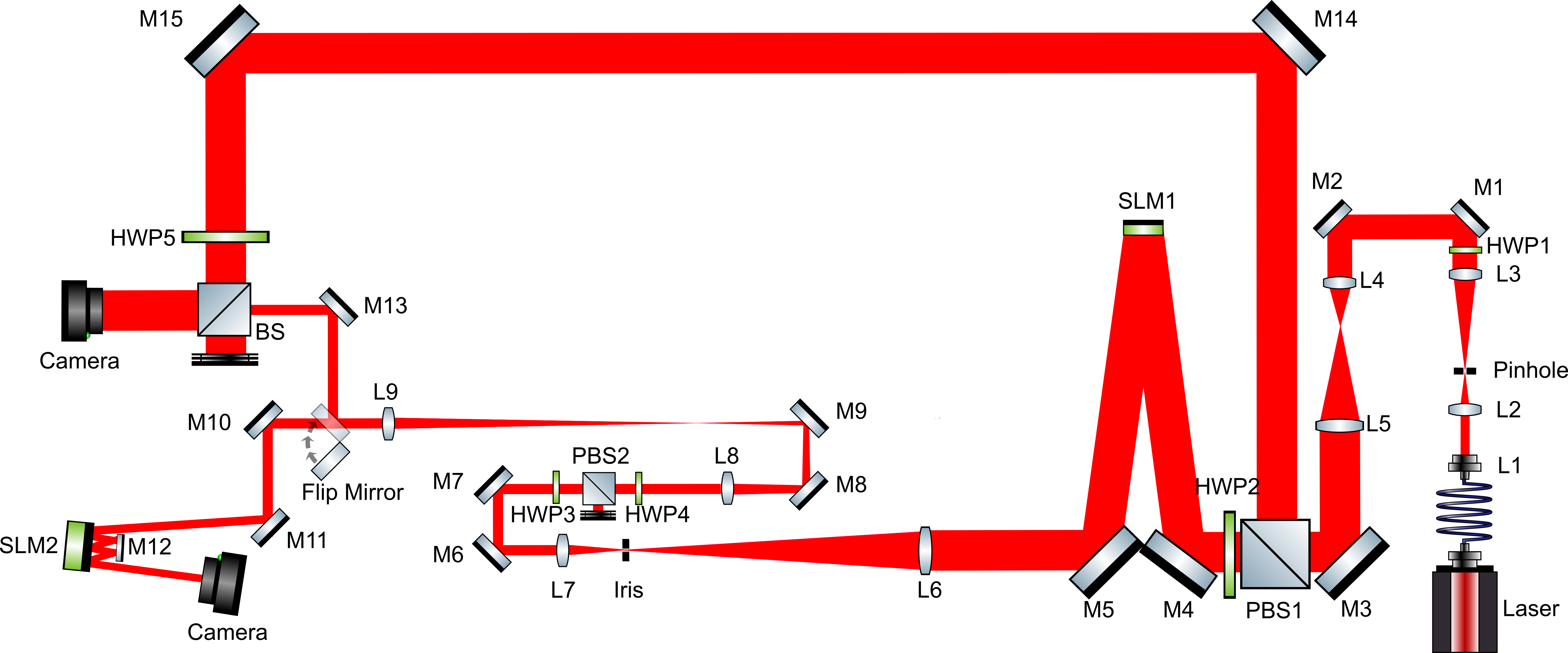}
\caption{\added{Full mode-generator and mode-sorting set-up including reference arm. }
}
\label{fig:FullExpSetup}
\end{figure*}

%this is in the next section but here for formatting
% \begin{figure*}[b]
% \centering\includegraphics[width=\textwidth]{Sorter-Draft/alignment-example.png}
% \caption{\added{Effect of shifting a single plate out of position for the loss function is a) solely focused on minimizing crosstalk ($\alpha = 0$) and b) solely focused on maximizing efficiency ($\alpha = 1$).}
% }
% \label{fig:Alignment}
% \end{figure*}

\added{
The laser (Toptica DL100 at 786nm) is coupled into a polarization maintaining fibre to clean the spatial mode and collimated at the output by L1, giving 22 mW of power. A pair of telescopes (L2, L3 and L4, L5) serve to expand the beam to cover the entire active surface area of SLM1 (Meadowlark 1920x1152). A 75 $\mu$m pinhole in the first telescope cleans the beam further. This SLM model works on 255 grey-level phase quantization across the 0--2$\pi$ range and modulates the vertical polarization. Half-wave plates and polarizing beam-splitters are used to ensure the beam is properly polarized. HWP1 decides the split between the reference arm for off-axis holography of HG modes (M14, M15) and the mode-generator. The modes are generated in the first order by SLM1. An iris in the focal plane of a telescope (L6, L7) selects this order, followed by an additional telescope (L8, L9). The flip mirror can be engaged for HG mode quality measurement, sending the generated mode to M13 to interfere with the reference arm (polarization adjusted with HWP6) on a beam-splitter. The resultant interference pattern is then imaged with a camera (UI-214 CCD). Without the flip mirror, the modes propagate (aligned by M10 and M11) to the mode-sorter consisting of multiple reflections between SLM2 (Meadowlark 1952x1152) and M12 which are spaced 4.25 cm apart. The output intensity is then recorded on a camera (UI-214 CCD) placed 14.3 cm after the last reflection from SLM2. To ensure an accurate measure of crosstalk, in view of a limited dynamic range of the camera, two images are taken at a high and low exposure. In the high-exposure image, saturated pixels are removed and the remaining are combined with the low-exposure image (normalizing by exposure time). These exposures are chosen dynamically during the experimental run to set the brightest possible image without saturation.}

\begin{table}[ht]
    \centering
    \caption{Lens Parameters}
    \label{tab:example}
    \begin{tabular}{lcc}
        \hline
        \textbf{Lens} & \textbf{Focal length (mm)} \\
        \hline
        L1 & 3  \\
        L2 & 50  \\
        L3 & 250  \\
        L4 & 100  \\
        L5 & 300  \\
        L6 & 500  \\
        L7 & 100  \\
        L8 & 400  \\
        L9 & 500  \\
        \hline
    \end{tabular}
\end{table}

\section{Fixed detection regions}
\label{appendix:SizingFairness}

\begin{figure*}[b]
\centering\includegraphics[width=\textwidth,height=0.5\textheight, keepaspectratio]{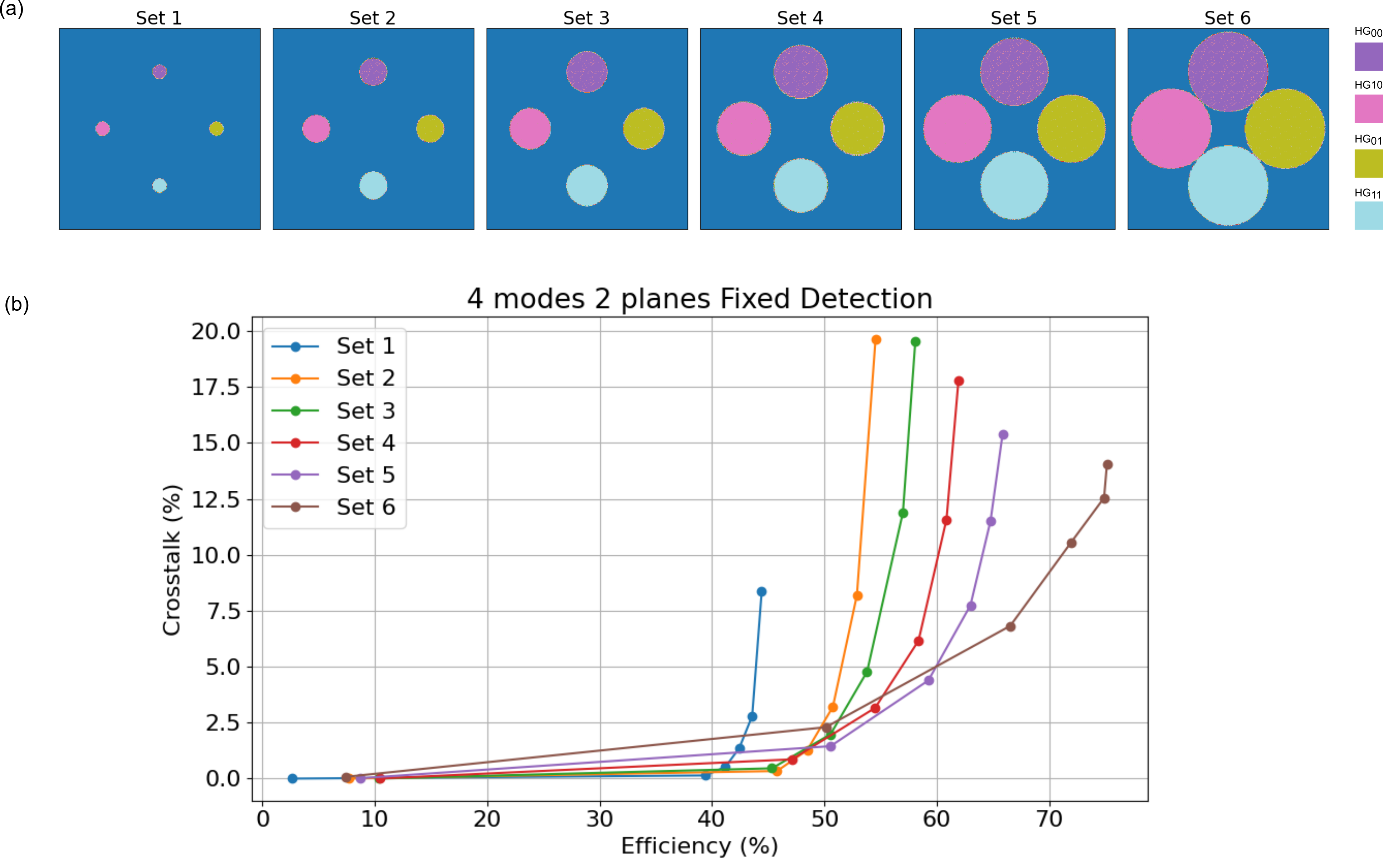}
\caption{\added{Effect of the detection region sizes (a) on efficiency and crosstalk performance (b). }
}
\label{fig:SizingFairness}
\end{figure*}

\added{ As stated in the main text, the fixed region sizes are chosen to balance efficiency and crosstalk. As Fig.~\ref{fig:SizingFairness} details, smaller more spatially separated detection regions will achieve lower crosstalks, while the larger less spatially separated detection regions achieve higher efficiencies. In order to balance these competing objectives and have a single fixed set of detection regions, we opt for a balanced option. For the case of 4 mode sorting this was Set 4 shown.} %further optimizing their size or spacing for either metric would effectively turn the baseline into a tunable detection rather than a fixed one.}

\section{Alignment Precision Requirements}
\label{appendix:Alignment}

\begin{figure*}[!b]
\centering\includegraphics[height=0.175\textheight, keepaspectratio]{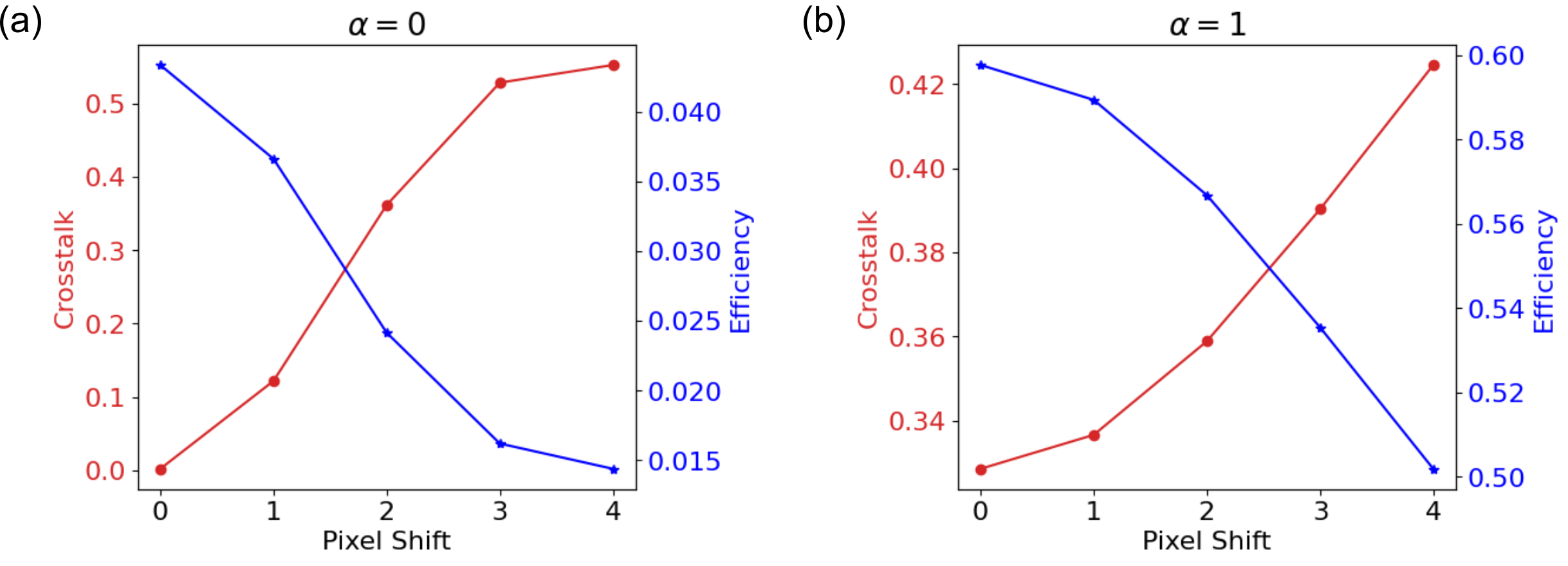}
\caption{\added{Effect of shifting a single plate out of position for the loss function is a) solely focused on minimizing crosstalk ($\alpha = 0$) and b) solely focused on maximizing efficiency ($\alpha = 1$).}
}
\label{fig:Alignment}
\end{figure*}

\added{To highlight the need for precise alignment we present here a simple example. Due to the symmetric nature of the flexible phase plates, without loss of generality we can consider offsets in the positive x direction, i.e. shifting the plate horizontally in its plane. For the case of a 3-plane sorter for 25 HG modes, we take the first plane and shift it out of position by a number of pixel steps, leaving the rest of the system fixed. The effect of this misalignment on the resulting crosstalk and efficiency is shown in Fig.~\ref{fig:Alignment}. One can see that the effects of misalignment are more severe for the case (a) of low crosstalk, low efficiency ($\alpha = 0$) than for the case (b) of high crosstalk, high efficiency ($\alpha = 1$).}

\end{document}